\documentclass[english,a4paper]{article}
\usepackage[T1]{fontenc}
\usepackage[cp1250]{inputenc}
\usepackage{amssymb}
\usepackage{hyperref}
\usepackage{amsfonts}
\usepackage{graphicx}

\begin{document}
\title{Fractional bidromy in the vibrational spectrum of HOCl}

\author{E. Ass\'emat, K. Efstathiou\footnote{Institute for Mathematics, University of Groningen, PO Box
  407, 9700AK Groningen, The Netherlands}, M. Joyeux\footnote{Laboratoire de Spectrom\'etrie Physique, CNRS UMR 5588,
  Universit\'e Joseph Fourier-Grenoble I, BP 87, F-38402 St Martin d'H\`eres
  Cedex, France} and D. Sugny\footnote{Laboratoire
  Interdisciplinaire Carnot de Bourgogne (ICB), UMR 5209 CNRS-Universit\'e de
  Bourgogne, 9 Av. A. Savary, BP 47 870, F-21078 DIJON Cedex, FRANCE, dominique.sugny@u-bourgogne.fr}}

\maketitle

\begin{abstract}
  We introduce the notion of fractional bidromy which is the combination of
  fractional monodromy and bidromy, two recent generalizations of Hamiltonian
  monodromy. We consider the vibrational spectrum of the HOCl molecule which is
  used as an illustrative example to show the presence of nontrivial fractional
  bidromy. To our knowledge, this is the first example of a molecular system
  where such a generalized monodromy is exhibited.
\end{abstract}


The description and the understanding of molecular spectra have been a
long-standing goal in the field of molecular physics, both from experimental and
theoretical points of view (for recent reviews, see \cite{joyeux,child} and
references therein).  Vibrational dynamics of sufficiently rigid polyatomic
molecules can be well reproduced up to a large fraction of the dissociation
threshold by an effective Hamiltonian which is obtained either by a fit of
parameters to a set of measured or calculated energy levels \cite{hcp}
or by the application of canonical perturbation theory to an ab initio potential
energy surface \cite{joyeux2}. An important class of effective Hamiltonians are
the classically integrable Hamiltonians, which allow to simplify the study of
the dynamics of the system by the use of constants of the motion.  Among them,
we can distinguish the simplest one, the Dunham expansion, which in the absence
of strong resonances describes accurately the vibrational dynamics at low energy
near a minimum of the potential energy surface. This effective Hamiltonian can
be written as a polynomial expansion in terms of the actions of the normal
modes, which can be defined globally on the whole phase space. Resonant
Hamiltonians, i.e., effective Hamiltonians with fundamental frequencies in
resonance \cite{joyeux}, describe the dynamics at higher energies where the coupling
between, at least, two degrees of freedom cannot be neglected. Even if
Hamiltonians of this kind are integrable, i.e., the number of constants of the
motion is equal to the number of degrees of freedom, their classical dynamics
may not be globally described by action-angle variables since the latter are in
general only locally defined. The question that naturally arises is how to
detect this feature both from the classical and quantum points of view. Indeed,
the fact that the actions are only local has a quantum counterpart in the joint
spectrum of the quantum Hamiltonian as it prevents the existence of global
quantum numbers \cite{san1}.

In this context, monodromy, which is the simplest topological obstruction to the
existence of a global set of action-angle variables, has become a useful tool
both in classical \cite{duist,cushman,efsbook} and quantum or semi-classical
mechanics \cite{san1,efsbook}. First discovered and developed by mathematicians,
the phenomenon of monodromy has been exhibited in a large variety of physical
systems extending from atomic and molecular ones \cite{sadov} to purely
classical systems \cite{cushman,dullin}. Such systems have a standard monodromy
which is either characterized by an isolated focus-focus singularity in the
associated bifurcation diagram for the local case or by a second leaf which is
glued to the main leaf through a line of bitori for the nonlocal situation
\cite{child}. Both types of monodromies appear in Fermi resonant
systems with a non zero angular momentum \cite{child2}. Recently, different
kinds of generalized monodromy, such as \emph{fractional monodromy} \cite{neko1}
and \emph{bidromy} \cite{bidromy1,bidromy2}, have been defined and their
presence shown in model Hamiltonian systems (see below for a concrete definition
of these generalizations). The next step in this study is the determination of
physical systems having such monodromies. It is in this spirit that we revisit
the analysis of the vibrational dynamics of the HOCl molecule with zero angular momentum. We show the
presence in this molecular spectrum of nontrivial fractional bidromy which can
be viewed as the combination of fractional monodromy and bidromy. We first
describe the corresponding bifurcation diagram, which presents a line of curled
tori and a line of bitori. Fractional bidromy is defined through a bipath, i.e.,
a set of two loops, which are allowed to cross both lines of curled tori and
bitori. This is a specificity of generalized monodromies with respect to
standard ones for which the associated loop lies in the set of regular values of the
bifurcation diagram. We determine the quantum monodromy matrix for a bipath such
that only one of its two components crosses the line of curled tori. Conclusion
and prospective views are given in the last section.\\
\emph{Vibrational dynamics of HOCl.}
Several studies have investigated the vibrational dynamics of HOCl both from the
experimental and theoretical points of view (details can be found in \cite{jost}
and references therein). In particular, although very accurate ab initio
calculations have been undertaken, it has been shown that the use of an
effective Hamiltonian allows an original and precise understanding of the
qualitative features of the dynamics \cite{jost}. This effective Hamiltonian
includes energy levels of the ground electronic state with an energy up to 98
$\%$ of the dissociation energy. The classical Hamiltonian $H$ is expressed in terms of the normal modes
coordinates $(q_1,p_1,q_2,p_2,q_3,p_3)$. $q_1$, $q_2$ and $q_3$ are close
respectively to the Jacobi coordinates $(r,\gamma,R)$, where $r$ is the OH bond
length, $R$ the distance from Cl to the center of mass G of OH ($R$ is very
close to the OCl bond length) and $\gamma$ the OGCl angle ($\gamma=0$ at linear HOCl
geometry). $H$ can be written as the sum of two terms $H=H_D+H_F$ where the
Dunham expansion $H_D$ and the resonant part $H_F$ are respectively equal to
\begin{eqnarray*} \label{eq2}
  H_D &=&\sum_{i=1}^3\frac{\omega_i}{2}(p_i^2+q_i^2)+\sum_{i\geq j}\frac{x_{ij}}{4}(p_i^2+q_i^2)(p_j^2+q_j^2)+ \nonumber \\
  && \sum_{i\geq j\geq k}\frac{y_{ijk}}{8}(p_i^2+q_i^2)(p_j^2+q_j^2)(p_k^2+q_k^2)+\cdots , \nonumber \\
  H_F&=&\frac{1}{\sqrt{2}}[(q_3^2-p_3^2)q_2+2q_3p_3p_2](k+\sum_i\frac{k_i}{2}(p_i^2+q_i^2)+ \nonumber \\
  &&\sum_{i\geq j}\frac{k_{ij}}{4}(p_i^2+q_i^2)(p_j^2+q_j^2)+\cdots).
\end{eqnarray*}
The parameters $\omega_i$, $x_{ij}$, $y_{ijk}$, $z_{ijkl}$, $k$, $k_i$ and $k_{ij}$
for HOCl can be found in \cite{jost}. Since the functions
$I_1=(p_1^2+q_1^2)/2$, $I=p_2^2+q_2^2+(p_3^2+q_3^2)/2$ and $H$ are constants of
the motion, the system is integrable. An additional 3:1 resonance between the
modes 1 and 2 can also be considered but we neglect it in this paper since it
renders the effective Hamiltonian non integrable. \\
\emph{Bifurcation diagram.}
Before describing the classical bifurcation diagram, we have to say some words
about the quantum problem. The quantization rules for nondegenerate vibrations
are \cite{jost}
\begin{eqnarray} \label{eq11}
  \left\{ \begin{array}{ll}
      I_1=\hbar(v_1+\frac{1}{2}) \\
      I=\hbar(P+\frac{3}{2})
    \end{array} \right. ,
\end{eqnarray}
where $v_1$ and $P=2v_2+v_3$ are respectively the number of quanta in the OH
stretching degree of freedom and the polyad number.  $v_1$ and $P$ are the
quantum numbers associated to the classical constants of the motion $I_1$ and
$I$. The constant $\hbar$ is an effective Planck constant which can, at least
theoretically, be modified at will to increase the density of energy
levels. This is necessary since the notion of quantum Hamiltonian monodromy is
only defined rigorously in the semi-classical limit where $\hbar\to 0$
\cite{san1}. The value $\hbar=1$ corresponds to the physical problem \cite{jost}.

\begin{figure}
  \includegraphics[width=\columnwidth]{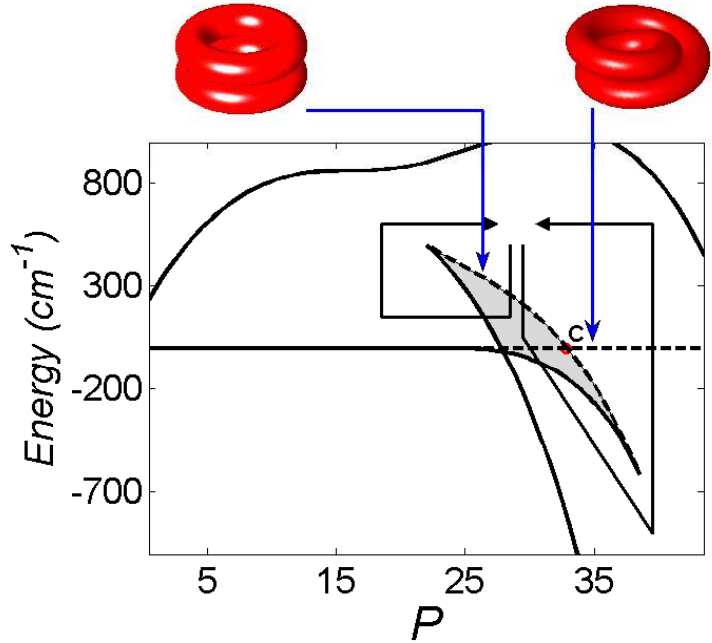}
  \caption{\label{fig1} (Color online) Bifurcation diagram of the HOCl molecule
    for $I_1=1/2$ as a function of $P$ [see Eq. (\ref{eq11})]. The energies of the
    period orbit [B] characterized by $p_3=q_3=0$ have been subtracted from each
    energy in order to clarify the plot (see Ref. \cite{jost} for details). The
    critical points of $\mathcal{EM}$ are represented by solid and dashed
    lines. Points of the solid lines lift to points or circles in the original
    phase space, while points of the dashed lines lift to curled tori or bitori. A
    representation of these two singular tori is also given in this figure. The red dot $C$ is the intersection point of these two lines.
    The two paths with arrows in black correspond to the bipath used to define the bidromy (see
    the text). The grey region is the zone where the two leaves of the bifurcation
    diagram overlap.}
\end{figure}

\begin{figure}[tbp]
  \includegraphics[width=\columnwidth]{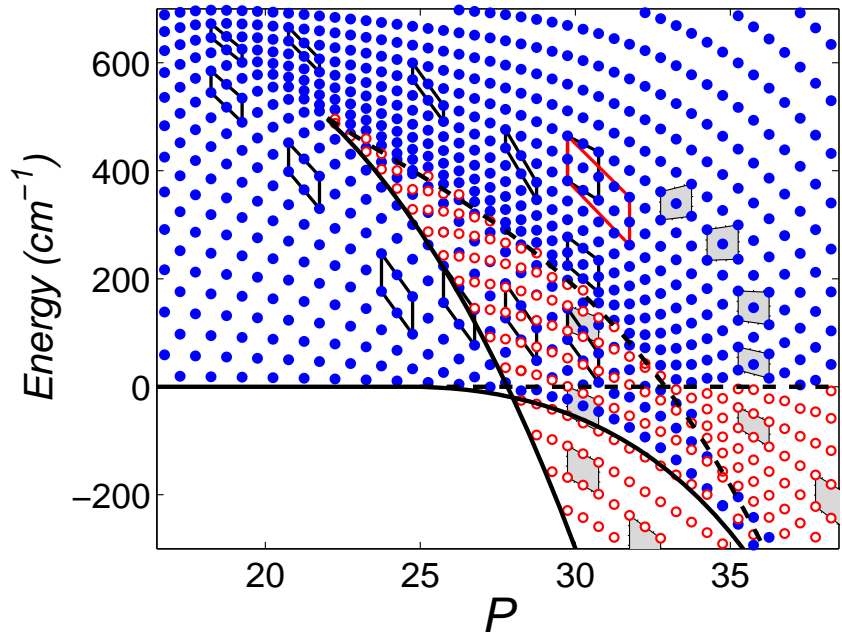}
  \caption{\label{fig2} (Color online) Parallel transport of the cell through
    the vibrational spectrum of HOCl. The energy levels of the upper and lower
    leaves are respectively depicted in blue (full) and red (open) dots. The
    effective value of $\hbar$ is taken to be 0.5 for the sake of a clearer illustration. The initial cell is broken into two cells when crossing the line of bitori. One of the two cells, the grey
    one, belongs to the lower leaf, while the other belongs to the upper
    leaf. The final cell, the red one, is defined by the addition of the two
    final cells of the two paths.}
\end{figure}

From now on, we only consider two degrees of freedom and we assume that there is
no excitation in the OH stretching, that is $v_1=0$ and $I_1=1/2$. The study
would be similar for other values of $I_1$. The bifurcation diagram is defined
as the image of the energy-momentum map $\mathcal{EM}$: $(q_2,p_2,q_3,p_3)\to (I=\mathcal{I},H=E)$.
This set is constructed from the determination of the critical values of
$\mathcal{EM}$ where the differentials $dI$ and $dH$ are not linearly
independent. Using the Liouville-Arnold theorem under suitable conditions, it
can be shown that the preimage of a regular (i.e., not critical) value of
$\mathcal{EM}$ is a two-dimensional torus. This also means that $\mathcal{EM}$
defines a torus-bundle with base space the regular values of the image of
$\mathcal{EM}$ and with generic fiber a torus. The preimage of critical values
is a critical fiber which can be of different types: point, circle, curled
torus, bitorus, etc. The topology of the corresponding critical fibers can be
determined from singular reduction theory \cite{cushman,efsbook}. The
bifurcation diagram of the HOCl molecule which has been derived in
Ref. \cite{jost} is displayed in Fig. \ref{fig1}. Note that the corresponding effective Hamiltonian is of degree larger than 4 and that this bifurcation diagram cannot be mapped on the catastroph map discussed in \cite{child,child2} for polyad numbers above the second bifurcation.
To better visualize the phenomenon of monodromy, we plot polyads up to
$P=44$. $P=38$ is the highest polyad number with an ab initio energy
level taken into account in the fit. As can be seen in Fig. \ref{fig1}, the
bifurcation diagram has a line of curled tori and a line of bitori which
intersect at the point $C$. It is also composed of two leaves, called the upper
leaf and the lower one, which overlap in the grey region of the bifurcation
diagram. The two leaves are glued together along the singular line of bitori and
along the part of the line of curled tori that lies at the right of the point
$C$. The other part of the line of curled tori belongs to the upper leaf (see also Fig. \ref{fig2} which displays the quantum version of this bifurcation diagram).

The vibrational energies can be obtained by a direct quantum computation in each
polyad, that is as a function of $P$. However, this calculation is not
sufficient to construct the quantum version of the bifurcation diagram and a
semi-classical analysis is needed to establish the nature of the classical
dynamics associated to each quantum energy level. For that purpose, we introduce
the canonical conjugate coordinates $(I_k,\phi_k)$ $(k\in\{1,2,3\})$ which are
defined by the relations $q_k=\sqrt{2I_k}\cos\phi_k$ and
$p_k=-\sqrt{2I_k}\sin\phi_k$. Note that the polar coordinates $(I_k,\phi_k)$ are not
defined if $p_k=q_k=0$. The Hamiltonian $H$ can then be expressed in the set of
coordinates $(I,\theta)$ and $(J,\psi)$ where $I=2I_2+I_3$, $J=2I_2$, $\theta=\phi_3$ and
$\psi=\phi_2/2-\phi_3$ with the constraints $I\geq J$ and $J\geq 0$. The Hamiltonian $H$
is a function of only $I$, $J$ and $\psi$. One of the actions of the system is $I$
which is global and the other one is given by $\mathcal{J}=\int_\gamma Jd\psi/(2\pi)$
where the integral is calculated along the projection $\gamma$ of the flow of $H$ on
the space $(J,\psi)$. $\mathcal{J}$ depends on the values of $I$ and $E$. The
regular Bohr-Sommerfeld rules state that the semi-classical energies are those
which satisfy $\mathcal{J}=\hbar(n+1/2)$ and $I=\hbar(P+3/2)$. Knowing the leaf to which
the loop $\gamma$ belongs, we associate the same leaf to the corresponding
semi-classical energy level. The accuracy of the semi-classical energy levels
allows us to do the same for each quantum energy level and to construct the
quantum bifurcation diagram. This diagram is displayed in Fig. \ref{fig2} where
we observe in the overlap region of the two leaves a superposition of two
lattices of points.\\
\emph{Fractional bidromy.}
Classical monodromy is the simplest topological obstruction to the existence of
global action-angle coordinates \cite{duist,cushman}.  Let us consider the
torus-bundle over the regular values of $\mathcal{EM}$. Due to the presence of
certain isolated singular tori such as pinched tori, regular tori are forced to
fit together with a twist which prevents extending the action-angle variables to
the whole bundle. The system has then a non trivial monodromy. From a quantum
point of view, we can also analyze the joint spectrum of the corresponding
quantum operators and look for the manifestation of classical monodromy in this
spectrum \cite{san1}. The bifurcation diagram becomes a 2-dimensional lattice of
points labeled by the values of the quantum numbers, the energy $E$ and the
polyad number $P$ for HOCl. Locally, around a regular value, the lattice is
regular in the sense that we can find a map which sends this lattice to
$\mathbb{Z}^2$. In order to check the global regularity of the spectrum, the method consists in
taking a cell, transporting continuously this cell along a closed loop and
comparing the final cell with the initial one. If the two cells are different
then the system has quantum (or at least semi-classical) monodromy \cite{san1}.

The theory of Hamiltonian monodromy has known recent important developments,
which resulted in the concepts of fractional monodromy \cite{neko1} and bidromy
\cite{bidromy1,bidromy2}. These generalizations are associated to particular
singular tori, i.e., curled tori for fractional monodromy and bitori for bidromy
(see Fig. \ref{fig1} for a representation of these singular tori), and to loops
in the bifurcation diagram which are allowed to cross these lines of
singularities. In \cite{neko1}, fractional monodromy characterizes a line of
critical values corresponding to curled tori and ended by a point whose preimage
is a pinched torus. If we consider loops crossing this line then it can be shown
that the notion of Hamiltonian monodromy can still be defined in a restrictive
way where the monodromy matrix has fractional coefficients. The bidromy
phenomenon was first introduced in a three-degree of freedom system similar to
the $\textrm{CO}_2$ molecule \cite{bidromy1} and exhibited recently in a general
class of two-degree of freedom Hamiltonian systems with a bifurcation diagram
having a swallowtail structure \cite{bidromy2} very close to the one encountered in
Fig. \ref{fig1}, except that there is in addition a line of curled tori in
Fig. \ref{fig1}. The bidromy matrix is defined through a bipath, i.e., a set of
two loops following each a different leaf of the bifurcation diagram. A bipath
appears as the only way to generalize the notion of monodromy when the
bifurcation diagram has a swallowtail structure and a line of bitori. Note the
difference between this case and the standard non local monodromy
\cite{efsbook,child} which is also characterized by a second leaf in the
bifurcation diagram but where loops surround the line of bitori without crossing
it. As for fractional monodromy, a method to cross unambiguously the line of
bitori can be defined. In the quantum version of these generalizations, the size
of the fundamental cell has to be increased as discussed below in one of the two
directions in order to parallel transport the cell across the line of critical
values.

Since the system considered in this paper has both a swallowtail structure and
a line of curled tori, we introduce the notion of fractional bidromy which can be viewed as the
combination of fractional monodromy and bidromy. As displayed in Fig. \ref{fig1}, we consider a bipath such that
one of its two components crosses the line of curled tori. In this example, we
restrict the determination of this generalized monodromy to the quantum case.
Let $u_1$ (resp. $v_1$) and $u_2$ (resp. $v_2$) be the two vectors defining the
initial (resp. final) cell. The quantum monodromy matrix $M$ is the matrix such
that
\begin{eqnarray}
  \left( \begin{array}{cc}
      v_1 \\ v_2 \end{array} \right)
  =M \left( \begin{array}{cc}
      u_1 \\ u_2 \end{array} \right).
\end{eqnarray}
The vectors $u_2$ and $v_2$ are vertical vectors oriented from the top to the
bottom, while the vectors $u_1$ and $v_1$ are oriented from left to right. To cross the different lines of singularities, the size of the cell has
to be increased in the horizontal direction for the line of curled tori
\cite{neko1} and in the vertical direction for the line of bitori
\cite{bidromy2}. The parallel transport of the cell along the bipath is
represented in Fig. \ref{fig2}. The cell is broken into two cells when crossing the line of bitori. The way this line is crossed is not a priori obvious. This parallel transport can be
rigorously computed by considering the classical monodromy and its relation with
quantum monodromy \cite{bidromy2}. After crossing this line of critical values,
the two cells are transported along the two parts of the bipath. We finally
merge the two final cells by adding their basis vectors, which gives the red
cell of Fig. \ref{fig2}. An analysis of Fig. \ref{fig2} allows to deduce the
following relations between the initial and final vectors: $v_1=2u_1+u_2/2$ and
$v_2=u_2$, which leads to the following monodromy matrix $M$:
\begin{eqnarray}\label{eqmatrix}
  M=\left( \begin{array}{cc}
      2 & 1/2 \\ 0 & 1 \end{array} \right).
\end{eqnarray}
Note that the initial vector $u_1'=u_1-u_2/2$ is transformed into $2u_1'$ after one loop, which leads in this case to a diagonal monodromy matrix. If we consider now a bipath crossing the line of bitori at the right of the
point $C$ then the same monodromy matrix is obtained although each component of
the bipath crosses two times the line of curled tori. This shows that the notion
of fractional bidromy is well defined as it does not depend on where the bipath
crosses the line of bitori.\\
\emph{Conclusion.}
This article proposes, to our knowledge, the first example of a molecular system
where a generalized monodromy is exhibited. We hope that this example will
motivate systematic investigations of generalized monodromy in the rovibronic
spectra of resonant molecular systems. Other two degree of freedom molecular systems such as HOBr \cite{azzam} are expected to present nontrivial generalized monodromies, but this situation is not general. As an example, the bidromy phenomenon does not exist in the HCP molecule \cite{hcp} since in this case one cannot define a closed bipath turning around the bifurcation points of the bifurcation diagram.
\bibliographystyle{apsrev}

\end{document}